\documentstyle[pre,aps,psfig]{revtex}
\begin{document}
\draft
\title{Towards a Simple Model of Compressible Alfv\'enic Turbulence} 
\author{M.V.~Medvedev \cite{email,mm} 
{ } and  P.H.~Diamond \cite{pd} }
\address{Physics Department, University of California at San Diego,
La Jolla, California 92093.}
\maketitle

\begin{abstract}
A simple model collisionless, dissipative, compressible MHD (Alfv\'enic) 
turbulence in a magnetized system is investigated. In contrast to more
familiar paradigms of turbulence,
dissipation arises from Landau damping, enters via nonlinearity, and is 
distributed over all  scales. The theory predicts that two different regimes
or phases of turbulence are possible,
depending on the ratio of steepening to damping coefficient $(m_1/m_2)$.
For strong damping ($|m_1/m_2|<1$), a regime of  smooth, hydrodynamic 
turbulence is predicted. For $|m_1/m_2|>1$, steady state 
turbulence does not exist in the hydrodynamic limit. Rather, spikey, 
small scale structure is predicted.  
\end{abstract}

The theory of compressible MHD (e.g., Alfv\'enic)  turbulence has been a 
topic of interest for some time \cite{turb}. Alfv\'en wave turbulence
presents several novel challenges, due to the fact the $k$--$\omega$
selection rules preclude three Alfv\'en-wave resonance. Thus, in 
incompressible MHD, two Alfv\'en waves can interact only with the vortex
(i.e., eddy) mode. Compressibility relaxes this constraint by allowing
interaction with accoustic and ion-ballistic modes (i.e., Landau damping),
along with waveform steepening. This naturally leads to the
formation of Alfv\'enic shocklets. Thus, one approach, which is analogous 
to the noisy-Burgers model in hydrodynamics, is
based on the study of nonlinear wave evolution equations with external noise 
drive [e.g., the noisy derivative nonlinear Schr\"odinger equation (DNLS) equation, 
in space physics].
Such theories describe turbulence as an ensemble 
of nonlinear structures, e.g., shocks, discontinuities, and high-amplitude 
waves,  which are typically observed in compressible (e.g., interplanetary 
\cite{disc}) plasmas.  This course of investigation was pursued 
computationally to study the noisy-DNLS equation \cite{nDNLS}. 
Stationarity was achieved by inserting {\em ad-hoc} viscous damping 
(later linked to finite plasma conductivity \cite{resist}) into the
otherwise  conservative DNLS equation. The DNLS model fails, however, 
for the important case of $\beta \sim 1$ ($\beta=4\pi p/B^2_0$ is the ratio 
of plasma pressure to magnetic pressure, $B_0^2$ is an external magnetic field)
and the electron-to-ion temperature ratio $T_e/T_i \sim 1$ 
(for instance, in the solar wind plasma), when Alfv\'en waves couple 
to strongly damped ion acoustic modes. As a consequence, the kinetically 
modified DNLS \cite{MD,all}, referred to as the kinetic nonlinear Schr\"odinger 
equation (KNLS), which exhibits intrinsically 
dissipative nonlinear coupling, emerges as the superior basic model. 
Numerical solution of the KNLS reveals a new 
class of dissipative structures, which appear through the balance of nonlinear 
steepening with collisionless nonlinear damping. 
These structures include arc-polarized and S polarized rotational 
discontinuities \cite{us-num}, observed in the solar wind plasma and
not predicted by other models. The resulting quasi-stationary structures
typically have narrow spectra.

Here, we present the first analytical study of 
the noisy-KNLS equation as a generic model of collisionless, large-amplitude
Alfv\'enic shocklet turbulence. Indeed, this is, to our knowledge, the
first structure-based theory of compressible MHD turbulence in a collisionless
system. Stationarity is maintained via the balance
of noise and dissipative nonlinearity. Dissipation here results from ion
Landau damping, which balances the parallel ponderomotive force 
produced by modulations of the compressible Alfv\'en wave train.
A one-loop renormalization group (RG)
calculation (equivalent \cite{DIA-RG} to a direct interaction approximation 
\cite{dia} closure) is utilized.  Although the KNLS describes both 
quasi-parallel and oblique waves \cite{MD}, 
we consider here the simpler case of quasi-parallel propagation. 
The general case will be addressed on future publication.
The noisy-KNLS is, thus, a generic model of strong, compressible
Alfv\'enic turbulence and may be relevant to the solar wind, interstellar 
medium, shock acceleration as well as to compressible MHD theory, as a 
whole. Note that this perspective is analogous to that of the noisy-Burgers 
equation model of compressible fluid turbulence \cite{nBurg}.
Several features which are
not common in standard MHD turbulence theories appear in this model. 
It is shown that the dissipative {\em integral} coupling renormalizes the 
{\em wave train velocity}, in addition to inducing nonlinear damping and 
dispersion. 
Moreover, consideration of the resulting solvability condition for a stationary
state in the hydrodynamic limit ($\omega,k\to0$) suggests that KNLS 
turbulence can exist in one of {\em two different states} or {\em phases}. 
In the hydrodynamic regime, turbulence consists of large-scale, smooth
($\omega,k\to0$) waveforms and dissipative structures. In the regime when
the hydrodynamic limit does not exist, one may expect a small-scale, spikey,
intermittent ($\omega,k\not\to0$) shocklet turbulence. 
This hypothesis, however, needs further (e.g., numerical) study.

The ``noisy-KNLS'' equation is
\begin{equation}
\frac{\partial\phi}{\partial t}+v_0\frac{\partial\phi}{\partial z}+
\lambda\frac{\partial}{\partial z}\left(\phi U_2\right)
-i\mu_0\frac{\partial^2\phi}{\partial z^2}=\tilde f ,
\label{KNLS}
\end{equation}
where $\phi=(b_x+ib_y)/B_0$ is the wave magnetic field, $\tilde f$ is the 
random  noise, $\mu_0= v_A^2/2\Omega_i$ is the dispersion coefficient,
$v_0$ is the reference frame velocity, $\lambda=1$ is the 
perturbation parameter, $v_A$ is the Alfv\'en speed, and  $\Omega_i$ is 
the ion gyrofrequency. Unlike the Burgers equation, the KNLS (and DNLS) 
equation is not Galilean
invariant, hence the $v_0$ term is explicit. The packet velocity $v_0$
is renormalized due the {\em broken symmetry} between $+k$ and $-k$ 
harmonics induced by Landau damping. This precludes the conventional
practice of transforming to the frame co-moving at $v_A$ to eliminate $v_0$.
The macroscopic  ponderomotive plasma velocity 
perturbation for a high amplitude Alfv\'en wave  is 
\begin{equation}
U_2=m_1|\phi|^2+m_2\widehat{\cal H}[|\phi|^2] , \qquad
\widehat{\cal H}=\frac{1}{\pi}\int_{-\infty}^\infty\frac{\cal P}{z'-z}
\textrm{d}z' ,
\end{equation}
where $\widehat{\cal H}$ is the (nonlocal) Hilbert operator which represents
collisionless (Landau) dissipation.
The coefficients $m_1$ and $m_2$ are functions of  
$\beta$ ($c_s$ and $T_e/T_i$, only, i.e.,
\begin{mathletters}
\begin{eqnarray}
m_1&=&\frac{1}{4}\frac{(1-\beta^*)+\chi_\|^2(1-\beta^*/\gamma)}{
(1-\beta^*)^2+\chi_\|^2(1-\beta^*/\gamma)^2} ,\\
m_2&=&-\frac{1}{4}\frac{\chi_\|\beta^*(\gamma-1)/\gamma}{
(1-\beta^*)^2+\chi_\|^2(1-\beta^*/\gamma)^2} ,
\end{eqnarray}
\end{mathletters}
where  $\beta^*=(T_e/T_i)\beta$, $\gamma=3$ is the polytropic constant,
and $\chi_\|=\sqrt{ 8\beta/\pi\gamma} \left[(T_e/T_i)^{3/2}
\exp{\{(T_i-T_e)/2T_i\} }\right]$ is the parallel heat conduction coefficient
which models kinetic collisionless dissipation in fluid models.
The term $m_1$ represents nonlinear steepening of a wave via coupling to
the self-generated density perturbation (associated with an acoustic mode). 
The term $m_2$
corresponds to {\em kinetic} damping of a wave by resonant particles, which
rapidly sinks energy from all harmonics, unlike viscous dissipation. We 
emphasize that the KNLS is intrinsically a {\em nonlinearly dissipative} 
equation, i.e. there is no {\em linear} damping retained here.
In this regard, we comment that there appear to be two meaningful paradigm 
problems to be explored. One is to retain {\em both} linear growth {\em and} 
damping in the KNLS.
In this case the results will necessarily be quite model dependent. The other,
which we pursue here, is to study the purely nonlinear problem with noisy drive.
This case allows us to isolate and focus on the intrinsically nonlinear dynamics
of the KNLS equation.

In Fourier space $\widehat{\cal H}=ik/|k|$, so the transformed KNLS is
\begin{eqnarray}
& &\left(-i\omega+iv_0k+i\mu_0k^2\right)\phi_{k\atop\omega}
\nonumber\\
& &{ }+i\lambda k\sum_{k', k''\atop\omega', \omega''}
\left(\phi_{k'\atop\omega'}\phi_{k''\atop\omega''}
\phi_{k-k'-k''\atop\omega-\omega'-\omega''}
\left[m_1+im_2\textrm{sign}(k-k')\right]\right)=f_{k\atop\omega} ,
\label{fourier}
\end{eqnarray}
where the function $\text{sign}(x)=x/|x|$. The stochastic noise 
$f_{k\atop\omega}$ is assumed to be zero-mean, and $\delta$-correlated in
space and time. To extract information from Eq.\ (\ref{fourier}), we 
utilize the direct interaction approximation (DIA) closure
\cite{dia,cubic,PD-Burg}. We follow the approach used in \cite{PD-Burg} which is
different from \cite{cubic}, where the correlation functions of cubic Schr\"odinger
turbulence were found. We emphasize the statistical nature of our analysis.
Indeed, turbulence renormalizes the coefficients of the evolution equation,
leaving its functional form unchanged. The renormalized phase velocity $v$
and dispersion $\mu$ (see discussion below) appear in the second order in $\lambda$.
Since the KNLS (and DNLS) is not Galilean invariant, the vertex $\lambda$ is 
also renormalized. This is a third order effect. Because of mathematical
difficulties, we exactly calculate $v$ and $\mu$, only, and for $\lambda$, we provide
a simple heuristic argument. In general, the noise finction $\tilde f$
is also renormalized at higher orders in $\lambda$. Usually, in one-loop RG
analyses, such turbulent corrections are assumed to be small.
Note that knowledge of turbulent $\lambda$ and $\tilde f$ is {\em not}
required for predicting the existence of the hydrodynamic limit (see below).

We expand $\phi_{k\atop\omega}$ in a power series with 
respect to the perturbation parameter $\lambda$:
$\phi_{k\atop\omega}=\phi_{k\atop\omega}^{(0)}
+\lambda\phi_{k\atop\omega}^{(1)}
+\lambda^2\phi_{k\atop\omega}^{(2)}+\cdots$ 
and equate terms, order by order, in $\lambda$.
To second order, we have
\begin{eqnarray}
& &\left(\omega-v_0k-\mu_0k^2\right)\phi_{k\atop\omega}^{(2)}
=\lambda k\sum_{k', k''\atop\omega', \omega''}
\phi_{-k'\atop-\omega'}^{(0)}\phi_{-k''\atop-\omega''}^{(0)}
\phi_{k+k'+k''\atop\omega+\omega'+\omega''}^{(1)}
\left[m_1+im_2\textrm{sign}(k+k')\right]
\nonumber\\
& &\qquad
=-i\lambda^2\sum_{k', k''\atop\omega', \omega''} k(k+k'+k'')\ 
G_0(k+k'+k'', \omega+\omega'+\omega'')
\nonumber\\
& &{ }\qquad\cdot
\left(\phi_{-k'\atop-\omega'}^{(0)}\phi_{k'\atop\omega'}^{(0)}\right)
\left(\phi_{-k''\atop-\omega''}^{(0)}\phi_{k''\atop\omega''}^{(0)}\right)
\phi_{k\atop\omega}^{(0)}
\left[m_1+im_2\textrm{sign}(k+k')\right]
\left[m_1+im_2\textrm{sign}(k+k'')\right] ,
\label{2nd}
\end{eqnarray}
where the bare propagator $G_0(\omega,k)=i/(\omega-kv_0-k^2\mu_0)$. In the DIA, 
we take $\phi_{k\atop\omega}^{(2)}\simeq\phi_{k\atop\omega}^{(0)}$.
The terms proportional to $k$ and $k^2$ in the left hand side act to 
modify $v_0$ and $\mu_0$. Thus, the nonlinear term of Eq.\ (\ref{2nd})
represents an amplitude dependent correction to {\em both the velocity and 
dispersion coefficients}, and Eq.\ (\ref{2nd}) is a recursive equation for
the renormalized 
coefficients $v$ and $\mu$. The fixed point of this recursion relation
gives the self-consistent values of these coefficients. 
Replacing the bare $v_0$, $\mu_0$ with their amplitude
dependent counterparts $v$, $\mu$, we write
\begin{eqnarray}
& &\left(\omega-vk-\mu k^2\right)=\frac{\lambda^2}{(2\pi)^4}
\int\!\!\!\int\!\!\!\int\!\!\!\int_{-\infty}^\infty d\omega'd\omega''dk'dk''\ 
\left|f_{k'\atop\omega'}\right|^2\left|f_{k''\atop\omega''}\right|^2
\nonumber\\
& &{ } \nonumber\\
& &{ } \cdot
\frac{k(k+k'+k'')}{\left|\omega'-vk'-\mu {k'}^2\right|^2
\left|\omega''-vk''-\mu {k''}^2\right|^2}\ 
\frac{\left[m_1+im_2\textrm{sign}(k+k')\right]
\left[m_1+im_2\textrm{sign}(k+k'')\right]}{\omega+\omega'+\omega''
-v\left(k+k'+k''\right)-\mu\left(k+k'+k''\right)^2} .
\label{main}
\end{eqnarray}
We should note that $v$ and $\mu$ will now assume complex values,
%
\begin{equation}
v=v_r+iv_i ,\qquad
\mu=\mu_r+i\mu_i .
\label{v,mu}
\end{equation}
The real parts, $v_r$ and $\mu_r$, represent the amplitude dependent speed
of a wave packet (note, there is a momentum transfer from waves to 
resonant particles in this model) and nonlinear dispersion (i.e. an amplitude 
dependent frequency shift in Fourier space), respectively.
The imaginary parts, $v_i$ and $\mu_i$, correspond to damping processes. 
In particular, $v_i$ describes the exponential damping ({\em \'a la} 
phase mixing of a wave packet) and $\mu_i$ describes turbulent, viscous 
dissipation. It is easily seen that for $m_2\to0$ the KNLS may be 
written in the co-moving frame with $v_0=0$. Thus, no additional 
phase-mixing terms appear, since the terms $k'k$ and $k''k$ vanish 
upon integration over $-\infty<\{k', k''\}<\infty$, in the hydrodynamic limit.
 The collisionless damping 
breaks this symmetry of the $+k$ and $-k$ parts of the spectrum, thus resulting
in the novel phase-mixing and phase velocity renormalization terms 
(analogous to nonlinear frequency shifts) encountered here.

We seek solutions in the hydrodynamic limit $\omega\to0$, $k\to0$. For
simplicity, we assume for noise the white noise statistics, 
i.e., $f_{k\atop\omega}=f$. This assumption
is not too artificial, since MHD waves are usually pumped at large scales
(small-$k$) and the large-$k$ tail is heavily damped by collisionless
dissipation, which is an increasing function of $k$.
Ordered by powers of $k$, the nonlinear term in the integrals  
contains $k^3, k^4,\dots$ contributions. However, the {\em hydrodynamic 
behavior} is completely determined by the small-$k,\omega$ limit. Thus, by
 omitting higher-$k$ terms, Eq.\ (\ref{main}) naturally splits into
two equation for $v$ and $\mu$, respectively. 
The $\omega', \omega''$-integrations
can be easily performed in complex plane. It is convenient to introduce 
dimensionless variables $x'=k'\tilde\mu/\tilde v,\ x''=k''\tilde\mu/\tilde v$.
The $k',k''$-integrals in Eq.\ (\ref{main}), diverge as $k',k''\to0$ 
(i.e. infrared divergence). The integrations can be performed consistently
only in the limit where the
infrared cut-offs satisfy the inequality $x'_c, x''_c\ll 1$.  Quite
lengthy, but straightforward complex integrations yield
\begin{mathletters}
\label{int-main}
\begin{eqnarray}
v_r+iv_i&=&-f^4\frac{\lambda^2}{(2\pi)^2}\frac{2\tilde m}{\tilde v^3}\Biggl\{
\frac{\tilde m}{2}\ln^2x_c+\tilde m\ln(1-2\bar\mu)+
\frac{\tilde m}{2}\left(\frac{\pi^2}{3}+3\right)\Biggr\} ,
\label{v}\\
\mu_r+i\mu_i&=&-f^4\frac{\lambda^2}{(2\pi)^2}\frac{\tilde m\tilde\mu}{\tilde v^4}
\Biggl\{\hat m\frac{\ln x_c}{x_c}-\frac{1}{x_c}\Bigl[\hat m\bar v
(\bar\mu-3)+\tilde m\bar v(5-3\bar\mu)\Bigr]
\nonumber\\
& &{ }-\ln^2x_c\tilde m\Bigl[ 1+2\bar v(1-\bar\mu)\Bigr]
+\ln x_c\Bigl[ 2\tilde m(\bar\mu-\bar v)
\nonumber\\
& &{ }-\hat m(4\bar\mu-
5\bar v+1)\Bigr] +F(\bar v,\bar\mu)\Biggr\}  ,
\label{mu}
\end{eqnarray}
\end{mathletters}
where $\tilde v=v-v^*, \tilde\mu=\mu-\mu^*$,
$m=m_1+im_2, \tilde m=[m-m^*]\, \text{sign}(v_i\mu_i), \hat m=m+m^*$,
$\bar v=v/\tilde v, \bar\mu=\mu/\tilde\mu$, and the 
dimensionless infrared cut-off is $x_c=k_{min}\mu_i/v_i$.   
The function $F(\bar v,\bar\mu)$ is positive definite and contains
no explicit divergences $\sim x_c$. Since we are concerned with the 
hydrodynamic limit, where $x_c\ll 1$, the detailed structure of this function 
is not significant. Returning to standard notation, extracting
real and imaginary parts, and keeping the leading, divergent (in $x_c$) terms,
we have from Eq.\ (\ref{v})
\begin{mathletters}
\begin{eqnarray}
v_rv_i^3&=&f^4\frac{\lambda^2}{(2\pi)^2}\frac{m_2^2}{2}\ 2\pi\sigma ,
\label{all-a}\\
v_i^4&=&f^4\frac{\lambda^2}{(2\pi)^2}\frac{m_2^2}{2}\
\ln^2\!\!\left(\frac{k_{min}\mu_i}{v_i}\right) ,
\label{all-b}
\end{eqnarray}
where $\sigma=\text{sign}(\mu_r/\mu_i)$.  As can be easily seen, 
$v_r/v_i\sim1/(\ln^2k_{min})\to 0$ as $k_{min}\to 0$, so  
we obtain from Eq.\ (\ref{mu})
\begin{eqnarray}
\mu_rv_i^3&=&f^4\frac{\lambda^2}{(2\pi)^2}\frac{m_2}{8k_{min}}
\Biggl(\ 4m_1\ln\!\!\left(\frac{k_{min}\mu_i}{v_i}\right)+3m_2
\frac{\mu_r}{\mu_i}\Biggr) ,
\label{all-c}\\
\mu_iv_i^3&=&f^4\frac{\lambda^2}{(2\pi)^2}\frac{m_2}{8k_{min}}\
m_1\frac{\mu_r}{\mu_i} .
\label{all-d}
\end{eqnarray}
\label{all}
\end{mathletters}
For the coefficients $v$ and $\mu$, we may now write 
\begin{equation}
\begin{array}{rl}
&\displaystyle{
v_r\sim v_i \ \text{sign}\!\!\left(\frac{\mu_r}{\mu_i}\right)\ln^{-2}x_c ,
\quad v_i\sim-f\sqrt{\lambda m_2 \ln x_c} ,}\\
&\displaystyle{
\mu_r\sim \mu_i\left(\frac{4m_1}{3m_2}\right) ,
\quad \mu_i\sim-\left(\frac{m_1}{m_2}\right)^2\frac{f}{k_{min}}
\sqrt{\frac{\lambda m_2}{\ln x_c}} .}
\end{array}
\label{scalings}
\end{equation}
Note that the factor $\ln x_c=\ln(k_{min}\mu_i/v_i)\sim\ln(\ln k_{min})$
makes an insignificant cut-off correction.
Nonzero $v_r$ arises due  to wave momentum loss via interaction with 
resonant particles and reflects the process whereby a nonlinear wave
accelerates in the direction of
steepening (i. e. $v_r>0$ for $\beta\leq1, T_e=T_i$), an effect which is observed in
numerical solutions of the KNLS equation \cite{us-num}. This effect is 
logarithmic for $k_{min}\to 0$. Negative $v_i$ corresponds to exponential
damping due to phase mixing, and is proportional to the dissipation rate $m_2$.
The coefficient $\mu_r$ represents turbulent dispersion, and the
coefficient $\mu_1<0$ corresponds to turbulent viscous damping.
By analogy with noisy-Burgers equation \cite{PD-Burg}, 
Eqs.\ (\ref{scalings}) for the turbulent transport coefficients yield the 
pulse propagation scaling exponents for the hydrodynamic regime, which are 
defined by divergences at the cut-off. For diffusion term, we have 
$\delta x^2/\delta t\sim\mu_i\sim|\delta x|~ (\sim1/k_{min})$, 
so that $|\delta x|\sim\delta t$. This corresponds to symmetric ballistic 
{\em dispersion} of the shocklet waveform. For the
velocity term, we write (as $v_r\to 0$ when $k_{min}\to 0$) 
$x/t\sim v_i\sim\sqrt{\lambda\ln x_c}\sim const$, that is
$x\sim t$. This corresponds to ballistic {\em translation} of the shocklet.

We now construct the quantity $x_c=k_{min}\mu_i/v_i$ from Eqs.\ (\ref{all}) 
to determine when our cut-off approximation $x_c\ll 1$ is valid. 
Note that $x_c\ll1$ must be satisfied for a self-consistent, 
hydrodynamic regime solution. Dividing Eqs.\ (\ref{all-b},\ref{all-c}) 
by Eq.\ (\ref{all-d}), we derive a system of equations which is easily 
simplified to give the condition
\begin{equation}
4x_c^2\ln^3x_c-3x_c\ln x_c-\left(\frac{m_1}{m_2}\right)^2=0 .
\label{x}
\end{equation}
Again, for $x_c\ll 1$, we may omit the small $x_c^2$ term. This equation
has maximum at $(x_c)_{max}=e^{-1}<1$, i.e. a solution of this equation 
for small $x_c$ exists only when $x_c\leq(x_c)_{max}$.
When $x_c>(x_c)_{max}$, Eq.\ (\ref{x})
does not have a small-$x_c$ solution, so no stationary state is possible in
the hydrodynamic limit.  To clarify the physical meaning of the
control parameter $x_c$, we write it as $x_c=(k^2_{min}\mu_i)/(k_{min}v_i)$.
Obviously, $x_c$ is just a measure of  the efficiency of turbulent {\em viscous} 
damping $\sim k^2$ relative to {\em collisionless} (Landau) damping 
(distributed in all scales). Smallness of 
$x_c$ indicates a situation of stronger Landau damping and weaker
linear turbulent (viscous) dissipation. The two cases of $x_c$ lesser or
greater $(x_c)_{max}$ thus correspond to different states of turbulence.
The regime of hydrodynamic turbulence [i. e. $x_c\leq (x_c)_{max}$] 
corresponds to strong damping, $|m_2|\gg|m_1|$, which dominates 
nonlinear steepening. Large-scale waveform structures are possible,
consistent with the notion of a hydrodynamic regime. The
turbulent viscous damping dominated nonlinear dispersion 
in this case, $\mu_r/\mu_i
\simeq 4m_1/3m_2\ll1$.
The opposite regime of ``shock'' turbulence (i. e. $x_c\geq (x_c)_{max}$) 
corresponds to weakly damped Alfv\'en waves, $|m_2|\ll|m_1|$
(however, Landau damping still dominates the small-scale dissipation), where
a stationary, hydrodynamic regime is not possible. In this case,
nonlinear steepening is balanced by turbulent dispersion, resulting in 
a state of small-scale coherent nonlinear structures, steep fronts and
discontinuities.
The bifurcation point can easily be found from $(x_c)_{max}\simeq e^{-1}$
and Eq.\ (\ref{x}) as
\begin{equation}
\left.\frac{|m_1|}{|m_2|}\right|_{bif}\simeq\sqrt{\frac{3}{e}}\simeq 1.1
\label{m1/m2}
\end{equation}
(the exact numerical solution yields $\simeq1.3$).
The coefficients $m_1, m_2$ depend on plasma parameters, i. e. on $\beta$
and $T_e/T_i$. We plot the condition Eq.\ (\ref{m1/m2}) in the form of a
$\beta$ vs. $T_e/T_i$ diagram in Fig.\ \ref{fig1}. The region inside the curve
corresponds to $|m_1/m_2|<1$, i.e. a phase of hydrodynamic turbulence. 
The outer region corresponds to a phase of bursty turbulence of steep 
nonlinear Alfv\'en waves.

For completeness, the perturbation parameter $\lambda$ must
be renormalized, because the KNLS is not Galilean
invariant. Corrections to $\lambda$ follow from the third order
expansion and so laborious that they are left for a future publication.
We can, however, estimate the renormalized $\lambda$
as follows. The energy spectrum is
\begin{eqnarray}
\tilde B^2(k)&=&\frac{1}{2\pi}\int_{-\infty}^\infty d\omega\
\frac{f_{k\atop\omega}f^*_{k\atop\omega}}
{\left|\omega-vk-\mu k^2\right|^2}
\nonumber\\
&=&-\frac{f^2}{2}\ \frac{1}{v_ik+\mu_ik^2} .
\end{eqnarray}
The fluctuation level
\begin{equation}
\tilde B^2=\frac{1}{2\pi}\int_{-\infty}^\infty dk\ \tilde B^2(k)
=-\frac{f^2}{4v_i}
\end{equation}
should be independent of the cut-off $x_c$, thus $\lambda\sim\ln^{-1}x_c$.
Of course, the fluctuation level may only depend on the noise strength, $f$, 
and the dissipation rate, $m_2$. As is expected, $\tilde B^2$ varies as 
$m_2^{-1}$ while noise is constant.

To conclude, we have presented the first analytical theory analysis
of a noisy-KNLS (and DNLS) model. The noisy-KNLS
describes turbulence of kinetically damped (at $\beta\simeq 1$) nonlinear
Alfv\'en wave turbulence, i.e. 
a turbulence of dissipative structures \cite{us-num}, discontinuities and
shock waves. The renormalized wave velocity and dispersion coefficients,
as well as the pulse propagation exponents, were calculated. 
Two different phases of turbulence were identified, depending on the
nonlinearity-to-dissipation coefficient ratio, $m_1/m_2$. For $|m_1/m_2|<1$
a stationary state of hydrodynamic ($k\to0, \omega\to0$)
turbulence (with noise) is predicted,
while for $|m_1/m_2|>1$ such a state is precluded and small-scale bursty, 
spikey turbulence is indicated. A phase diagram in the space of $\beta$
and $T_e/T_i$ is 
given. These findings may be pertinent to recent observations of multiple
states in solar wind plasma turbulence.

We thank B.~Tsurutani, V.D.~Shapiro, V.I.~Shevchenko, and S.K.~Ride
for useful discussions.
This work was supported by DoE grant No. DEFG0388ER53275, 
NASA grant No. NAGW-2418, and NSF grant No. ATM 9396158.

%
\begin{figure}
\caption{The $\beta$-$T_e/T_i$-diagram of state.
The region inside the curve corresponds to highly damped turbulence. No
steep fronts appear. There is wave steepening in the region outside the curve. }
\label{fig1}
\end{figure}
\newpage
\sf
\large\sf
\vspace*{2in}
\hspace*{3.8in}
\makebox(0,0){\psfig{file=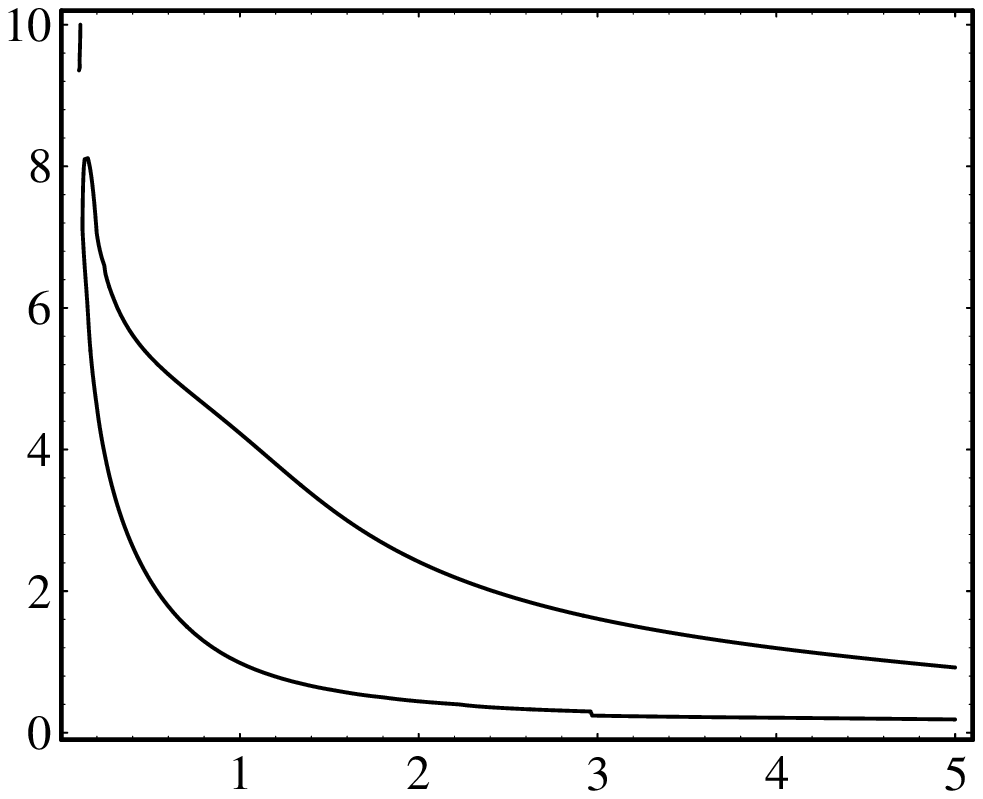,height=8in,width=7in}}
\put(-25,51){$\beta$}
\put(-194,180){$\displaystyle{\frac{T_e}{T_i}}$}
\put(-105,95){hydrodynamic}
\put(-155,85){bursty}
\put(-30,140){bursty}
\thispagestyle{empty}
\end{document}